\documentclass[12pt,preprint]{aastex}
\newcommand{\Msun}{\ifmmode\mbox{M}_{\odot}\else$\mbox{M}_{\odot}$\fi}
\newcommand{\Rsun}{\ifmmode\mbox{R}_{\odot}\else$\mbox{R}_{\odot}$\fi}
\newcommand{\Mearth}{\ifmmode\mbox{M}_{\oplus}\else$\mbox{M}_{\oplus}$\fi}
\newcommand{\Rearth}{\ifmmode\mbox{R}_{\oplus}\else$\mbox{R}_{\oplus}$\fi}
\newcommand{\chandra}{\textit{Chandra}\,}
\newcommand{\Suzaku}{\textit{Suzaku}\,}

\newcommand{\swift}{\textit{Swift}\,}
\newcommand{\xmm}{\textit{XMM}\,}
\newcommand{\asca}{\textit{ASCA}\,}
\newcommand{\xte}{\textit{RXTE}\,}

\newcommand{\seven}{RXS~J170849.0$-$400910\,}
\newcommand{\etfo}{1E~1841$-$045\,}

\newcommand{\ttfn}{1E~2259$+$586}
\newcommand{\tfe}{1E~1048.1$-$5937\,}

\shorttitle{Search for X-ray Variability in 1E~1841$-$045}
\shortauthors{Zhu, Kaspi}

\begin{document}
\title{Searching for X-ray Variability in the Glitching Anomalous X-ray
Pulsar 1E~1841$-$045 in Kes 73}
\author{Weiwei Zhu\altaffilmark{1},
Victoria M. Kaspi\altaffilmark{1,2},
\altaffiltext{1}{\footnotesize Department of Physics,
McGill University, Montreal, QC, H3A 2T8, Canada;
zhuww@physics.mcgill.ca, vkaspi@physics.mcgill.ca }
\altaffiltext{2}{Canada Research Chair; Lorne Trottier Chair;
  R. Howard Webster Fellow of CIFAR}
}
\begin{abstract}
Anomalous X-ray Pulsars (AXPs) are now established to exhibit significant X-ray
variability and be prolific glitchers, with some
glitches being accompanied by large radiative changes.
An open issue is whether AXP glitches are generically accompanied by radiative
changes, relevant for understanding magnetar physical properties.
Here we report on an analysis of archival X-ray data from the
AXP 1E~1841$-$045, obtained between 1993 and 2007.  This AXP, located in the
center of SNR Kes~73, has exhibited three glitches between 2002 and 2007, as
determined by {\it RXTE} monitoring since 1999.  We have searched for evidence
of phase-averaged flux variability 
that could be present if glitches in AXPs are usually accompanied by radiative
changes.  We find no evidence for glitch-correlated flux changes from
this source, arguing that such behavior is not generic to AXPs.
\end{abstract}

\keywords{pulsars: individual (1E 1841$-$045) --- X-rays: stars --- stars: neutron}

\section{Introduction}
\label{sec:intr}

X-ray variability in AXPs is now established as a characeteristic property of this source
class.  Their fluxes and spectra vary on a variety of different time scales,
from a few milliseconds (e.g. \citealt{gkw02}), to several years
\citep{dkg07}.  See \citet{kas07,mer08} for recent reviews of AXP variability.

AXPs are also now known to be prolific glitchers, and include some of the most
active glitchers known in the neutron-star population \citep{dkg08}.  By `glitch'
we mean a sudden rotational spin-up, or similar timing anomaly.  One
of the largest glitches seen so far, from AXP \ttfn, was accompanied by
major radiative changes, including bursts and a factor of $\sim$20 pulsed and
persistent flux increase \citep{kgw+03,wkt+04}.
\citet{dkg09} showed that glitches in AXP 1E~1048.1$-$5937 were also accompanied
with radiative enhancements.
Such events
are thought to be the result of sudden yielding of the neutron-star crust
due to internal stresses caused by the decay of the magnetar-strength field.
The restructuring results in changes to the stellar interior -- as evidenced
by the glitch -- and to the stellar exterior -- as evidenced by the dramatic radiative
changes.  Observations of such AXP outbursts are a potentially powerful probe of
the physics of magnetars \citep[e.g.][]{es10}.

Links between the X-ray variability
and glitches of AXP \seven\ have also been reported \citep{roz+05,cri+07,igz+07,gri+07}.  
These authors suggest the existence of a general correlation between
magnetars' flux and glitch epochs.  
Specifically, \citet{gri+07} reported $\sim$40\% flux changes in \seven\
based on eight observations made over the course of 
$\sim$9~yr, during which they report four glitches.
If correct, this suggests that 
glitches are usually, and possibly always, accompanied with radiative changes.
However, the sparsity of observing epochs compared to the number of
glitches for this AXP thus far is problematic in proving the variability is
glitch-correlated.  Additionally, no comparable pulsed flux changes were
observed in the same time span, during which such measurements were available
regularly on a monthly basis \citep{dkg08}.  This apparent conflict
could, however, be explained if the pulsed fraction were precisely
anti-correlated with flux.  An anti-correlation between pulsed fraction
and flux has been seen in AXP 1E~1048.1$-$5937 \citep{tmt+05,tgd+08}, although
not to a degree that render pulsed flux variations absent.

Here we investigate the hypothesis that AXP glitches are generically accompanied
by radiative changes by considering AXP \etfo  in supernova remnant (SNR) Kes
73, one of the most
frequent glitchers among AXPs.  From {\it RXTE} monitoring, we know that
\etfo has had three glitches between 1999
and 2008 \citep{dkg08}. The glitches occurred on 2002 July 9, 2003 December 24,
and 2006 March 29, and had $\Delta\nu/\nu $ of
$5.63\times10^{-6}$, $2.45\times10^{-6}$ and $1.39\times10^{-7}$, respectively,
and were not accompanied by any X-ray pulsed flux changes.
If, as seen in \ttfn\ and 1E~1048.1$-$5937 and reported for \seven, glitches are generically
accompanied by radiative changes, and, if as for \seven, such radiative
changes are not necessarily apparent in the pulsed flux data, it is possible that
the phase-averaged flux (unavailable from \xte monitoring) varies in concert with glitches in \etfo.
We investigate this possibility here.  Also worth noting is that
\citet{gvd99} studied the timing of AXP \etfo using archival {\it GINGA}, \asca,
{\it ROSAT} and \xte data taken between 1993 and 1999 and found no glitches with
$\Delta\nu/\nu > 5\times 10^{-6}$. 

In this paper, we report on our analysis of archival X-ray
data for \etfo\ collected by {\it ASCA, Chandra, XMM} and \Suzaku\ during the past 17 years,
including two observations that were made fortuitously very closely following glitches.
We have looked for correlations between the AXP's flux variability
and glitch epochs. In section \ref{sec:obs} we describe the observations and
data reduction process. In section \ref{sec:spec} we describe our spectral
analysis and AXP flux extraction method.  Our results and conclusions are
discussed in section \ref{sec:sum}.

\section{Observations}
\label{sec:obs}

For this study, we searched online X-ray archives for all existing observations
of the Kes~73 field.  We found a total of eleven observations from four different
focussing X-ray observatories (listed in Table~1).
We do not include in our analysis the many {\it RXTE} observations, as
due to its non-focussing nature, 
these provide only pulsed flux measurements, and are already published \citep{dkg08}.
Next we report on our analysis of the eleven focussing-telescope observations.

\subsection{\asca\ Observations}

Seven \asca \citep{tih94} observations of the AXP \etfo\, and the SNR Kes 73 were taken between 1993 and 1999 (see Table \ref{obstab}). 
Our analysis began with the screened data from the two Gas Imaging Spectrometers (GISs,
\citealt{ascagis}), which were filtered with the standard revision 2
screening criteria.
Given the angular resolution of the GISs, the SNR ($\sim$2$'$) was unresolvable
in the images, therefore the spectra we extracted from the GISs contain 
photons from both the AXP and the SNR.
Using the {\tt ftool
xselect}\footnote{http://heasarc.gsfc.nasa.gov/ftools/xselect/}, we extracted spectra from source regions of 9$'$.8 radius 
(a region large enough to encircle the extended emission from AXP \etfo and Kes 73)  
and background spectra from regions of radius $\sim$5$'$, away from
the source region for all the \asca GIS observations. 
The GIS spectra were then combined with Redistribution Matrices File (RMF)
of the GISs and Auxiliary Response File (ARF)
files generated using the {\tt ftool} {\tt ascaarf}, and grouped with a
minimum of 25 counts per bin.
Finally the exposure of the grouped spectra were corrected for the deadtime
effect using the {\tt ftool deadtime}.
In this study, we did not include the spectra extracted from the two
Solid-state Imaging Spectrometers (SISs, \citealt{ascasis}), primarily because of the
significantly fewer counts collected by these instruments.

\subsection{\chandra\ Observation}
\label{sec:chandra}

The AXP and SNR were observed by the \chandra\ {\it X-ray Observatory}
with the Advanced CCD Imaging Spectrometer (ACIS, \citealt{gbf+03}) in timed
exposure (TE) mode on 2000 July 23 and in continuous clocking (CC) mode on
2000 July 29 (Table \ref{obstab}).
The data were analyzed and reported by \citet{msk+03c}.
Here we used {\tt Ciao} version 3.4\footnote{http://131.142.185.90/ciao3.4/index.html}.
Because the spectrum of the pulsar in TE mode was heavily affected by
pile-up, we did not use these data; 
instead we extracted the spectrum of AXP \etfo from the CC mode data.
In these data, the image of the pulsar and SNR were collapsed into one dimension.
From the level 2 event list provided by the \chandra X-ray Center, we
extracted the spectrum of the pulsar using a box-shaped region capturing a 2$''$.5 long segment along
the one-dimensional image and centered on the pulsar. 
The background spectrum was extracted from two 7$''$.5 long segments adjacent
to the source region.
The resulting source and background spectra were then combined with RMF and
ARF files generated using the {\tt psextract} command and
grouped with a minimum of 25 counts per bin.

\subsection{\xmm\ Observations}
\label{sec:xmm}

The AXP was observed by the {\it Newton X-ray Multi-Mirror Mission} (\xmm)
observatory \citep{jla+01} on 2002 October 5 and 2002 October 7 (Table
\ref{obstab}) with the European Photon Imaging Camera (EPIC) pn
\citep{sbd+01} camera operating in large window mode and the EPIC MOS
cameras \citep{taa+01} in full window mode. 
For our analysis, we used
the \xmm Science Analysis System ({\tt SAS}) version 8.0.0\footnote{See http://xmm.esac.esa.int/sas/8.0.0/} and 
calibrations (updated 2008 Oct 3).
Given the pn and MOS cameras' angular resolution, \etfo\, can be resolved
from Kes 73.
For the two \xmm observations, we used only the data from the EPIC pn camera to
take advantage of its larger photon collecting area and to avoid
cross-calibration issues between the pn camera and the mos cameras.
We extracted the pulsar's spectrum from a circular region of radius 32$''$.5
(a radius large enough to capture more than 90\% of the photon events from the point source)
centered on the pulsar.
Background spectra were extracted from an annular region of radius between
35$''$ and 115$''$ centered on the pulsar.
The pulsar spectrum was grouped with a minimum of 25 counts per bin and then
combined with the background spectrum and RMF and ARF files generated 
by the {\tt SAS} software.

\subsection{\Suzaku\ Observations}

AXP \etfo and Kes 73 were also observed by the \Suzaku observatory \citep{suzaku} on 2006 April
19 (Table \ref{obstab}). 
The data analysis was reported by
\citet{mks+08}. Onboard \Suzaku, there are two
X-ray detectors: the X-ray Imaging Spectrometer (XIS, \citealt{suzakuxis},
consisting of four CCDs sensitive in soft-X-ray band) and the Hard X-ray Detector (HXD,
\citealt{suzakuhxd}, sensitive to 10--600 keV X-rays).
Here we present a spectral analysis of the XIS data only.
Given the angular resolution of the XIS, the SNR was unresolvable in the XIS
image.  Therefore the spectra we extracted from the XIS detectors contain 
photons from both the AXP and the SNR.
We used cleaned events screened by the standard pipeline processing version
2.0.6.13\footnote{http://heasarc.nasa.gov/docs/suzaku/processing/criteria\_xis.html}. 
The source spectra were extracted from a circular region of 260$''$ radius.
Background spectra were extracted from an annulus region of radius between 260$''$ and 520$''$.
The extracted spectra were grouped with a minimum of 25 counts per bin, and
then combined with the RMF and ARF files generated using the {\tt ftools xisrmfgen} and {\tt xissimarfgen}.

%


\section{Spectroscopy}
\label{sec:spec}
Among the eleven observations we used, the AXP can
be resolved from the SNR Kes 73 only in the \chandra\ and 
\xmm observations.  For these,
it is possible to extract either the neutron star's spectrum or the
combined spectrum of the neutron star and the SNR. 
By contrast, only combined spectra can be extracted from the \asca and \Suzaku observations.

The spectra of AXPs are often parametrized by a blackbody plus a power-law (although
this is known to be an approximation to a likely Comptonized blackbody spectrum --
see Thompson et al. 2002 and, for example, Rea et al. 2008).  \nocite{tlk02,rzt+08}
The spectra of SNRs are often fit with models like, for example, VSEDOV
(a plane-parallel shock radiation model with separate
ion and electron temperatures), VNEI 
(a non-equilibrium ionization collisional plasma model),
or VPSHOCK (a plane-parallel shocked plasma model). See \citealt{blr01} for a
review of these models.
In this paper, we modeled the neutron star
radiation with a blackbody plus power-law (BB+POW), and the SNR radiation with a VSEDOV
model, using {\tt xspec
12.5.0}\footnote{http://heasarc.nasa.gov/docs/xanadu/xspec/}.
The focus of our investigation is on the AXP's X-ray flux; in modelling the SNR we
sought only a suitable parameterization to allow us to subtract off
its flux reliably.  As we show below, the VSEDOV model is adequate for
these purposes.
We modeled the interstellar absorption by multiplying a WABS (a photo-electric absorption model, in which the interstellar absorption is
characterized by a single parameter $N_H$, the neutral hydrogen column density along the line of
sight) model to both the BB+POW and VSEDOV models.

\begin{figure}
\includegraphics[scale=0.6,angle=270]{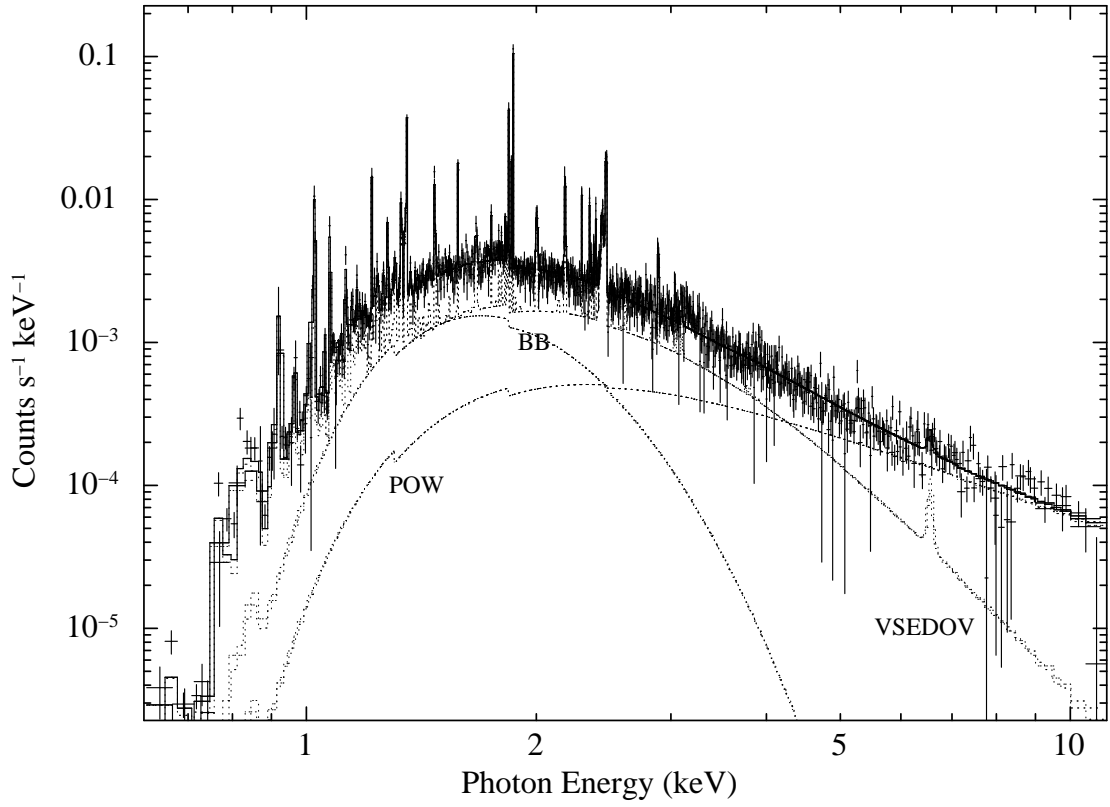} \\
\caption { \label{fig:spec} The unfolded \xmm pn spectra and components of
the best-fit BB+POW+VSEDOV model. 
}
\end{figure}

Figure \ref{fig:spec} shows the combined spectra from \xmm and the components
of the best-fit model. 
The AXP power-law component clearly dominates the spectra above $\sim$4
keV. 
Therefore, we chose to study the AXP flux in the 4--10~keV band only, in order
to minimize SNR contamination. 
Nevertheless,
we still attempted to remove the remaining small contribution of SNR in this band for the fluxes measured
from \asca and \Suzaku observations, so that we could compare them with the
neutron-star-only fluxes measured with \chandra and \textit{XMM}.

It is reasonable to assume that both the interstellar absorption and SNR
radiation do not change over a time scale of about a decade.
Consequently, in our attempt to remove the SNR flux, we used the same $N_H$
and VSEDOV parameters for all the spectra of the different observations, and
allowed only the BB+POW model to vary from observation to observation.

When fitting a WABS(BB+POW+VSEDOV) model to those
spectra containing emission from both the AXP and SNR, it is challenging to
constrain the normalization parameters of both the BB
and VSEDOV models simultaneously, because these two models dominate the same
energy band and their parameters are highly covariant.
Fortunately, the neutron star can be spatially resolved out in 
the \xmm observations, so we can use them to determine the relative strength of the two
spectral components.
We therefore fit a
WABS(BB+POW+VSEDOV) model to the combined (AXP+SNR) \xmm\ spectrum.
To ensure that we had the correct BB+POW model for the neutron star, we simultaneously
fit the spectrum extracted from only the neutron star, requiring common neutron-star
spectral parameters.  Hence we could determine the parameters and normalization 
of the VSEDOV model for the SNR.

To further improve the spectral model, next we included the 
\asca\ and \Suzaku\ spectra and the \chandra\ CC-model neutron-star spectrum
and performed a large joint fit, which required multiple iterations
in order to converge.
The result was a value for $N_H$ and for the 
VSEDOV model parameters that fit all the SNR spectra reasonably well, albeit not perfectly
(reduced $\chi^2 = 1.36$ for 3801 degrees of freedom for the joint fit).
The best-fit $N_H$ and VSEDOV parameters can be found in Table
\ref{tab:SNRmodel}.

Note that for all the spectral fitting described above, we used the
0.8--10 keV band.  However for \Suzaku, we found that
there were always significant residuals in the range 1.7--3.5 keV and above 9
keV, and in general, these residuals differed significantly among the four XIS instruments.
Therefore,
we ignored the 1.7--3.5 keV band and above 9 keV for the \Suzaku spectra.
Furthermore, the SNR Kes 73 is larger than the field-of-view of the \Suzaku
XISs and was not entirely captured.
Therefore, when fitting the \Suzaku\ spectrum, we allowed the VSEDOV normalization parameter to vary.

Finally, by using the best-fit $N_H$ and VSEDOV model, we could remove the
SNR flux contribution from the \asca and \Suzaku observations in separate
fits to their spectra, hence measuring the AXP fluxes. 
For the \chandra and \xmm observations, we simply fitted the resolved AXP
spectra with an absorbed BB+POW model.
The best-fit BB+POW parameters and the measured 4--10 keV unabsorbed fluxes of
the AXP are presented in Table \ref{tab:AXPmodel}.
The fluxes are also plotted in Figure \ref{fig:flux}.
The uncertainties we report on the 4-10 keV unabsorbed AXP fluxes were estimated by 
the measured fractional uncertainties on the 4--10 keV absorbed total
fluxes. 
For those observations in which the best-fit reduced $\chi^2$ was larger than
unity, we multiplied the reported flux uncertainties by 
the square root of the reduced $\chi^2$. 
This is to account for systematic errors in our imperfect modelling of the SNR spectrum.

To confirm the robustness of the measured AXP flux values,
we repeated the entire above analysis by modelling the SNR spectra
in different ways.
We found that both the VNEI and VPSHOCK models can fit the SNR
spectra as well as the VSEDOV model, and that the 4--10 keV unabsorbed neutron star
fluxes we measured using these models are consistent within reported uncertainties with the values
we found using the VSEDOV model. 
Therefore, we feel confident that the phase-averaged AXP flux values we report
are well constrained and robust.
However, the BB+POW spectral parameters we measured when fitting the SNR
with different models were not as robust as the 4--10 keV fluxes, changing
significantly with SNR model.  As we do not believe them to be reliably determined,
we do not quote their uncertainties in Table \ref{tab:AXPmodel},
and do not consider them further.

Also in Figure \ref{fig:flux}, we show the AXP's 2--10~keV pulsed flux 
as measured in monitoring observations with \xte since early 1999. 
Details about how these pulsed fluxes were determined are provided in
\citet{dkg08}.  The {\it RXTE} pulsed fluxes show no significant variations.
The vertical lines in Figure \ref{fig:flux} indicate the epochs of the three
observed glitches \citep{dkg08}.

\section{Discussion and Summary}
\label{sec:sum}

The goal of this study was to see whether the prolific glitching 
AXP 1E~1841$-$045 shows 
phase-averaged flux variability, in spite of showing no evidence for
pulsed flux variability.  Also, we wished to determine whether any variability is
correlated with its glitches as has been seen in AXP \ttfn\ in its
2002 major outburst, in 1E~1048.1$-$5937, and also reported for \seven.

\begin{figure}
\includegraphics[scale=1.0]{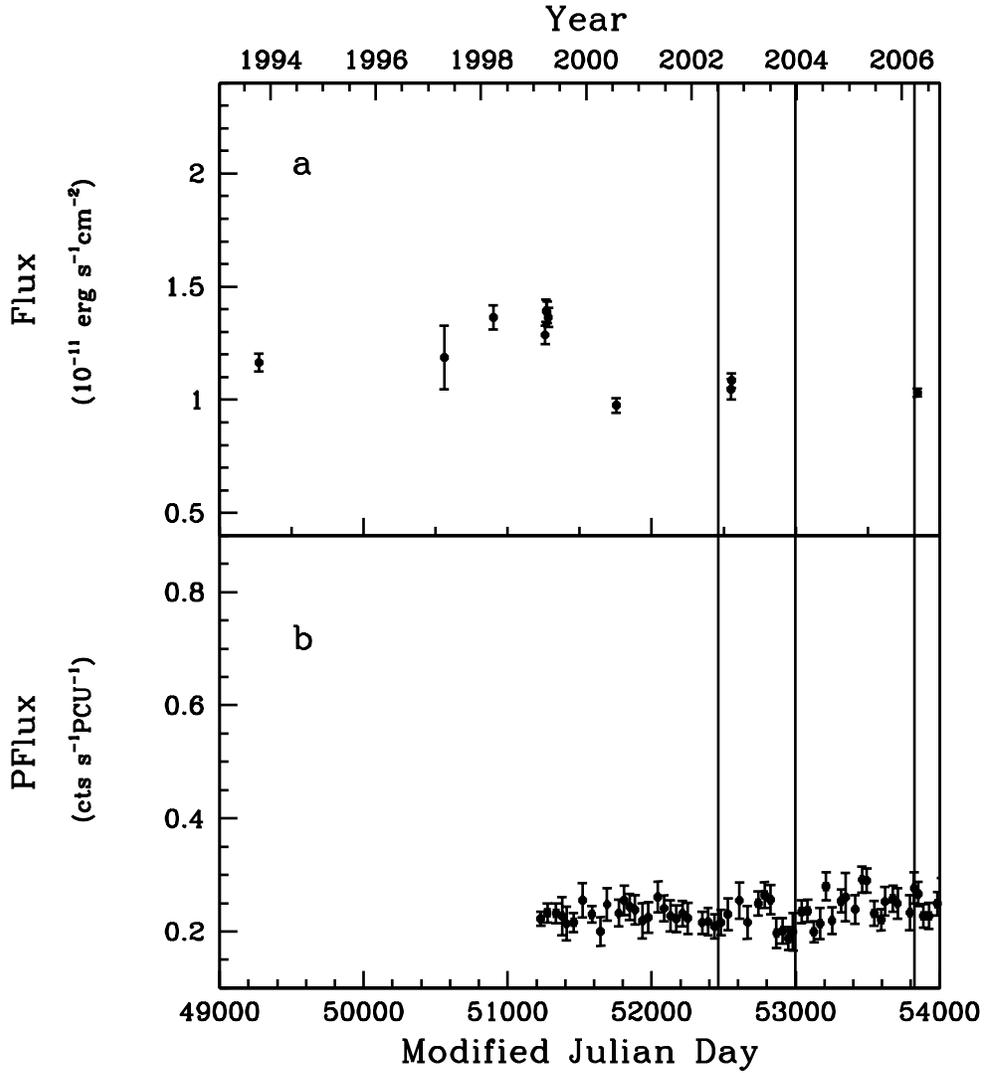} \\
\caption { \label{fig:flux}
(a) 4--10 keV unabsorbed phase-averaged fluxes from AXP 1E~1841$-$045, as determined by our analysis. (b) 2--10 keV
pulsed fluxes from the pulsar as measured by \xte, in units
of counts per second per Proportional Counter Unit (PCU)  \citep[see][for details]{dkg08}.
}
\end{figure}

As is clear from Panel a of Figure \ref{fig:flux}, in the 4--10 keV band, the
neutron star's flux did not vary by more than $\sim$30\% in 13 years.  Interestingly
the largest variations we find are in the multiple pre-1999 {\it ASCA}
observations:  in those seven observations, a fit to a constant flux 
results in a reduced $\chi^2$ of 3.7 for 6 degrees of freedom, which
has a probability of occurring by chance of $\sim$0.001.
However, during this time, there were certainly no large 
glitches ($\Delta \nu/\nu < 5\times 10^{-6}$) \citep{gvd99}.

The \asca fluxes all appear to be higher than the fluxes measured from other
observations. 
If this difference is accurate, then the AXP's flux must have dropped around
2000, and we might in principle expect some change in the 
pulsed flux as monitored by \xte\ at that epoch. 
However, 
the pulsed flux of the pulsar was constant around that time as
seen in Panel b of Figure \ref{fig:flux}.
Therefore, we suspect that the relative increase in the flux in the {\it ASCA}
observations compared with those of the other observatories may be
due to instrumental calibration issues.

The fluxes measured from the last four observations taken by
\chandra, \xmm and \Suzaku can be fitted with a constant flux model
(reduced $\chi^2 = 1.9$ for 3 degrees of freedom, corresponding to a probability
of having occurred by chance of 0.125).
Thus we conclude that the phase-averaged 4--10~keV fluxes of the last four observations were 
consistent with being constant, and we put an upper limit of 11\% on long-term
variability in this energy band.
Importantly, the two \xmm observations were taken only 88 and 90 days after the
first glitch, and the \Suzaku observation was taken only 27 days after the third
glitch.   
By contrast, the 4--10 keV flux of \ttfn\, was 50\% higher than in quiescence 21 days after its 2002
glitch \citep{zkd+08}, that of \tfe was a factor of 6 higher $\ge$38 days after
its 2007 glitches \citep{tgd+08}, and was 50--70\% higher for \seven\,
$\sim$53 days after its first 2005 glitch as inferred from
\citet{gri+07}\footnote{This is calculated based on the reported 1--10 keV
fluxes of \seven\, from its 2003 \xmm observation and 2005 \swift observation. 
We assumed a power-law to blackbody flux ratio of 3 for the 1-10 keV fluxes,
and then calculated the 4--10 keV fluxes using
webPIMMS (http://heasarc.gsfc.nasa.gov/Tools/w3pimms.html)}.
Therefore, we conclude that unlike in \ttfn, 1E~1048.1$-$5937 and possibly \seven, there is no
evidence for glitch-correlated flux changes in AXP \etfo.

One caveat of our study is that we were limited to the harder part of the
neutron star's emission spectrum.  The flux from the blackbody component 
was not well constrained.  Therefore, we cannot rule out changes in the 
neutron star's thermal radiation, only changes in the power-law
component in the 4--10~keV band, which we note constitutes
$\sim$0.25 of the stellar flux (BB+POW) in the 1--10 keV band.

Thus our results argue against the hypothesis that there exists a generic
correlation between the X-ray flux variability and glitch epochs in AXPs, 
further supporting the argument that glitches in AXPs can be either
radiatively loud or radiatively silent \citep{dkg08}.
There is of course precedent for radiatively silent glitches in neutron stars, in that
no rotation-powered pulsar glitch has ever been reported to be accompanied with
any radiative change, although rapid X-ray follow-up has only been accomplished
in one case \citep{hgh01}.
Any physical model of magnetar glitches will have to explain the simultaneous
existence of both types. This is true of even a single source, as there is
evidence that AXP \ttfn has both, given that its most recent glitch showed no
pulsed flux change \citep{dkg08b}.
Recently \citet{es10} have argued that AXP glitches are triggered
by energy releases at depths below $\sim$100~m in the crust, with angular
momentum vortex unpinning being due to global mechanical motion
triggered by the energy release, not by heat as has been proposed
in the context of rotation-powered pulsars glitches \citep{le96,lc02}.  If mechanical
triggering occurs, then
radiatively silent glitches of the same amplitude as radiatively loud glitches 
are possible since less energy is required
to trigger a glitch than to cause a substantial X-ray brightening.
If so, then \citet{es10} predict that all AXP glitches should occur simultaneously with or before the
observed X-ray brightening; this can be tested with continuous (daily or
better) X-ray monitoring observations.  Moreover, although mechanical unpinning of vortices by activity
in the lower crust does result in heat release, the latter could take as much as several years
to reach the surface.  This could help explain the long-term X-ray variability trends
that have been reported in some AXPs \citep[e.g.][]{dkg07}.

\acknowledgements
We thank Koji Mukai of the \asca GOF for assistant with the \asca data and David Eichler for useful conversations.  We also
thank Rim Dib for providing the {\it RXTE} pulsed fluxes.
VMK acknowledges support from NSERC via a Discovery Grant, FQRNT via
the Centre de Recherche en Astrophysique du Qu\'ebec, CIFAR and holds
a Canada Research Chair and Lorne Trottier Chair in
Astrophysics and Cosmology.

\bibliographystyle{apj}
\bibliography{myrefs,journals1,modrefs,psrrefs,crossrefs}

\clearpage
\begin{deluxetable}{lccc}
\tabletypesize{\footnotesize}
\tablewidth{0pt}
\tablecaption{\label{obstab} X-ray observations of AXP \etfo used in
this study. }%
\tablehead{ \colhead{Date} &\colhead{Observatory} &
\colhead{$t$\tablenotemark{a} (ks)} & \colhead{Offset\tablenotemark{b}
($'$)} }
\startdata
1993 Oct 11& \asca& 40& 6.8        \\
1997 Apr 21& \asca& 9& 7.3         \\
1998 Mar 27& \asca& 39& 6.5        \\
1999 Mar 22& \asca& 20& 5.3        \\
1999 Mar 29& \asca& 20& 5.2        \\
1999 Apr 06& \asca& 19& 5.2        \\
1999 Apr 13& \asca& 21& 5.2        \\
2000 Jul 29& \chandra& 10& 0.097   \\
2002 Oct 05& \xmm& 3.8& 1.152      \\
2002 Oct 07& \xmm& 4.4& 1.144      \\
2006 Apr 19& \Suzaku& 98& 3.9      \\
\enddata
\tablenotetext{a}{The effective exposure time of the instrument used for
the spectral analysis in this paper.}
\tablenotetext{b}{The pointing offsets of the observations relative to the
position of the AXP.}
\end{deluxetable}

\clearpage
\begin{deluxetable}{lc}
\tabletypesize{\footnotesize}
\tablewidth{0pt}
\tablecaption{\label{tab:SNRmodel} WABS and VSEDOV parameter values }
\tablehead{ \colhead{Parameter}  &\colhead{Value}   }
\startdata
$N_H$&  2.77 $\times10^{22}$cm$^{-2}$\\
$kT_a$&  0.56 keV\\
$kT_b$&  0.56 keV\\
$Mg$&  0.63\tablenotemark{a}\\
$Si$&  1.10\\
$S$&  1.92\\
$Ca$&  0.32\\
$Fe$&  0.57\\
$Ni$&  2.51\\
$\tau$&  $1.50\times10^{11}$ s\\
Redshift&  0.00\\
Norm\tablenotemark{b}&  0.33\\
$H$...$Ar$\tablenotemark{c}&  1.0
\enddata
\tablenotetext{a}{The element abundances quoted here are the relative
abundances based on the Solar mixture abundances.}
\tablenotetext{b}{Normalization parameter for the VSEDOV model, $\frac{10^{-14}}{4\pi[D_A(1+z)]^2\int n_e n_H dV}$ where $D_A$ is the angular diameter distance to the source (cm), and $n_e,n_H$ (cm$^{-3}$) are the electron and hydrogen densities respectively.}
\tablenotetext{c}{Elements $H$...$Ar$: $H$ $He$ $C$ $N$ $O$ $Ne$ and $Ar$ were fixed to the solar abundance 
because the SNR spectra are not sensitive to them. }

\end{deluxetable}

\clearpage
\begin{deluxetable}{cccccc}
\tabletypesize{\footnotesize}
\tablewidth{0pt}
\tablecaption{\label{tab:AXPmodel} Measured AXP model parameters and fluxes. }
\tablehead{ \colhead{Date}  &\colhead{Observatory}  &\colhead{MJD}  &\colhead{$kT$\tablenotemark{a} (keV)}  &\colhead{$\Gamma$\tablenotemark{a}}  &\colhead{$F_{NS}$\tablenotemark{b} (10$^{-11}$erg cm$^{-2}$s$^{-1}$)}   }
\startdata
1993 Oct 12&  \textit{ASCA}&  49272.1&  0.20&  2.72&  1.16(4)\\
1997 Apr 21&  \textit{ASCA}&  50559.6&  0.19&  3.00&  1.2(1)\\
1998 Mar 27&  \textit{ASCA}&  50899.95&  0.20&  2.70&  1.36(5)\\
1999 Mar 22&  \textit{ASCA}&  51259.7&  0.20&  2.89&  1.29(4)\\
1999 Mar 30&  \textit{ASCA}&  51267.0&  0.19&  2.80&  1.39(5)\\
1999 Apr 06&  \textit{ASCA}&  51274.6&  0.21&  2.74&  1.39(5)\\
1999 Apr 14&  \textit{ASCA}&  51282.3&  0.21&  2.68&  1.36(4)\\
2000 Jul 29&  \textit{Chandra}&  51754.32&  0.40&  2.10&  0.975(3)\\
2002 Oct 05&  \textit{XMM}&  52552.16&  0.22&  2.59&  1.05(4)\\
2002 Oct 07&  \textit{XMM}&  52554.16&  0.21&  2.47&  1.08(3)\\
2006 Apr 19&  \textit{Suzaku}&  53844.8&  0.38&  1.94&  1.03(2)
\enddata
\tablenotetext{a}{The reported best-fit $kT$ and $\Gamma$ parameters of the BB+POW model vary depending on the 
assumed SNR model; we do not report the fit uncertainties of these parameters as
they do not reflect the true uncertainties.  The values, determined while
assuming the VSEDOV model for SNR, are provided for reference only.}
\tablenotetext{b}{Phase-averaged unabsorbed neutron-star flux in the 4--10 keV band. The number in parenthesis represents the 1$\sigma$
uncertainties in the last digit. See text for details.}

\end{deluxetable}

\clearpage

\clearpage

\end{document}